\documentclass[
    ,final            
    ,numberedheadings 
  ,bibliocites
  ]
  {aipproc}

\layoutstyle{6x9}

\usepackage{natbib}

\begin{document}
\title{On the Detectability of the Cosmic Dark Ages: 21-cm Lines from Minihalos}


\author{Hugo Martel}{
address={Department of Astronomy, University of Texas at Austin}
}
\author{Paul R. Shapiro}{
address={Department of Astronomy, University of Texas at Austin}
}

\author{Ilian T. Iliev}{
address={Osservatorio Astrofisico di Arcetri, Italy}
}

\author{Evan Scannapieco}{
address={Osservatorio Astrofisico di Arcetri, Italy}
}

\author{Andrea Ferrara}{
address={Osservatorio Astrofisico di Arcetri, Italy}
}


\def\gtrsim{\mathrel{\raise.3ex\hbox{$>$}\mkern-14mu
             \lower0.6ex\hbox{$\sim$}}}
\def\lesssim{\mathrel{\raise.3ex\hbox{$<$}\mkern-14mu
             \lower0.6ex\hbox{$\sim$}}}

\begin{abstract}
In the standard Cold Dark Matter (CDM) theory of
structure formation, virialized minihalos (with 
$T_{\rm vir}\le10,000{\rm K}$) form in abundance at
high redshift ($z\gtrsim6$), during the cosmic ``dark
ages.'' The hydrogen in these minihalos, the first
nonlinear baryonic structures to form in the
universe, is mostly neutral and sufficiently hot
and dense to emit strongly at the 21-cm line. We
calculate the emission from individual minihalos
and the radiation background contributed by their
combined effect. Minihalos create a ``21-cm forest''
of emission lines. We predict that the angular 
fluctuations in this 21-cm background should be
detectable with the planned LOFAR and SKA radio
arrays, thus providing a direct probe of structure
formation during the ``dark ages.'' Such a detection
will serve to confirm the basic CDM paradigm while
constraining the background cosmology parameters,
the shape of the power-spectrum of primordial
density fluctuations, the onset and duration of
the reionization epoch, and the conditions which
led to the first stars and quasars. We present
results here for the currently-favored, flat $\Lambda$CDM
model, for different tilts of the primordial power
spectrum. These minihalos will also cause a ``21-cm forest'' of
absorption lines, as well, in the spectrum of
radio continuum sources at high redshift, if the
latter came into existence before the end of
reionization.
\end{abstract}
\maketitle

\section*{Introduction}

No direct observation of the universe during the period between the
recombination epoch at redshift $z\simeq10^3$ and the reionization epoch
at $z\gtrsim6$ as yet been reported. While a number of suggestions for the
future detection of the reionization epoch, itself, have been made,
this period prior to the formation of the first stars and quasars --
the cosmic ``dark ages'' -- has been more elusive. Standard Big-Bang 
cosmology in the CDM model predicts that nonlinear baryonic structure 
first emerges during this period, with virialized halos of dark and
baryonic matter which span a range of masses from less than $10^4M_\odot$
to about $10^8M_\odot$ which are filled with neutral hydrogen atoms. The
atomic density $n_{\rm H}$ and kinetic temperature $T_K$ of the gas
are high enough that collisions populate the hyperfine levels of the ground
state of these atoms in a ratio close to that of their statistical weight
(3:1), with a spin temperature $T_S$ that greatly exceeds the
excitation temperature $T_*=0.0681{\rm K}$. Since
$T_S>T_{\rm CMB}$, the temperature of the Cosmic Microwave Background 
(CMB), as well, for the majority of halos, these ``minihalos'' can be 
a detectable source of redshifted 21-cm emission. The direct detection 
of minihalos at such high redshift would be an unprecedented measure
of the density fluctuations in the baryons and of the total matter power 
spectrum at small scales.
The results presented here are described in more detail in
\cite{isms02,isfm02,si03}.

\section*{21-CM EMISSION AND ABSORPTION}

The 21-cm emission from a single halo depends upon its internal atomic
density, temperature, and velocity structure. We model each CDM minihalo here
as a nonsingular, truncated isothermal sphere (``TIS'') of dark matter 
and baryons in virial and hydrostatic equilibrium, in good agreement with the results of gas and N-body simulations from realistic initial conditions
\cite{is01,is02,sir99}. 
The minihalos which
contribute significantly to the 21-cm emission span a mass range from
$M_{\min}$ to $M_{\max}$ which varies with redshift. $M_{\min}$ is close to
the Jeans mass $M_J$ of the uncollapsed IGM prior to reionization,
while $M_{\max}$ is the mass for which $T_{\rm vir}=10^4{\rm K}$
according to the TIS model \cite{is01}. Typically, at $z=9$,
$M_{\min}\sim\hbox{few}\times10^3M_\odot$ 
and $M_{\max}\sim\hbox{few}\times10^7M_\odot$,
depending upon the cosmological model.

\begin{figure}[b!] 
  \includegraphics[height=2.7in]{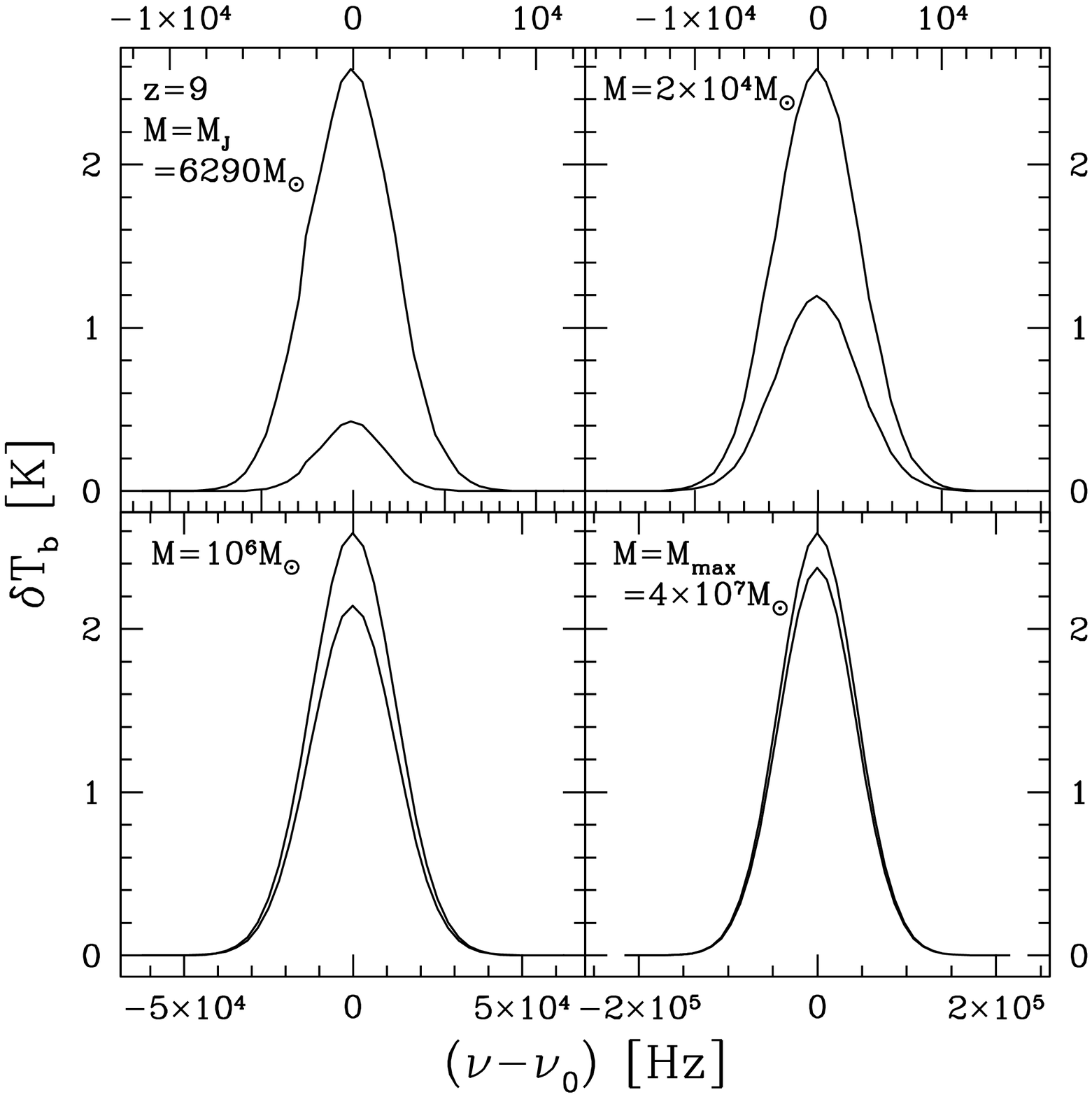}
  \includegraphics[height=2.7in]{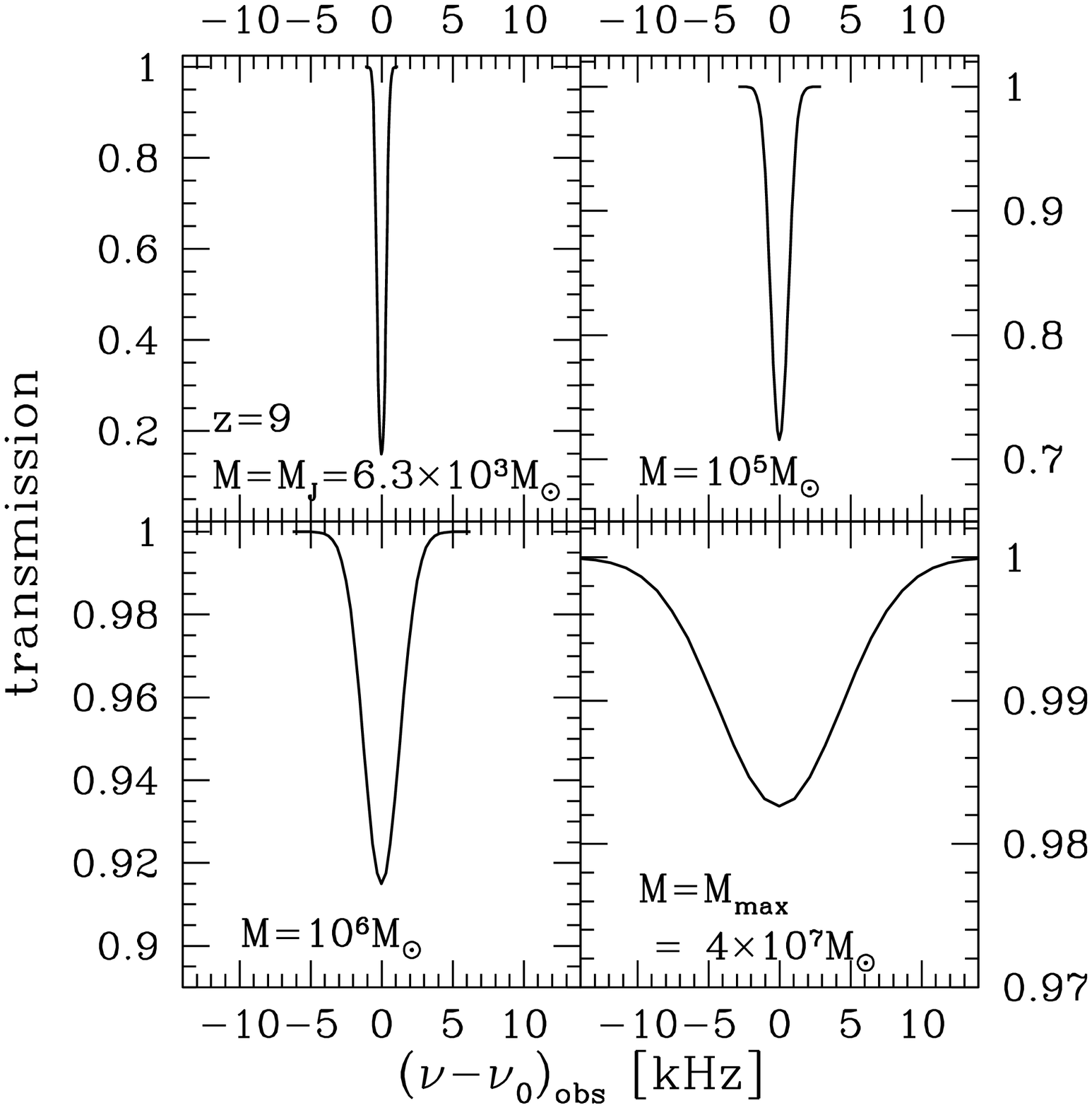}
\vspace{10pt}
\caption{(a) (left) 21-cm emission line profiles for individual minihalos of
different mass at $z=9$. Differential antenna temperature 
$\overline{\delta T_b}[{\rm K}]$
versus emitted frequency $\nu_{\rm em}$,
from detailed radiative transfer calculation (lower curves) and optically
thin approximation (upper curves). Optical depth is important, 
particularly for smaller halos.
(b) (right) 21-cm absorption line profiles for individual minihalos of different
mass at $z=9$. Transmission factor versus received frequency $\nu_{\rm rec}$
at $z=0$ for line of sight through minihalo center (i.e. zero 
impact parameter).}
\label{Martel_ea}
\end{figure}

Our results for individual minihalos are summarized in
Figure~\ref{Martel_ea}a. Line profiles of different minihalos along the same
line of sight should not typically overlap.  
The proper mean free path for
photons to encounter minihalos in $\Lambda$CDM is 160 kpc at $z=9$
\cite{s01}, corresponding to a frequency separation 
$\Delta\nu_{\rm sep}\sim
0.1\,{\rm MHz}\gg\Delta\nu_{\rm eff}\lesssim10\,{\rm kHz}$. These results 
predict a ``21-cm forest'' of minihalo emission lines. At $z=9$, for 
example, there are about 160 lines per unit redshift along a
typical line of sight\cite{s01}.
Detecting the stronger lines would require sub-arcsecond spatial resolution,
$\sim1\,{\rm kHz}$ frequency resolution, and $\sim$nJy sensitivity. SKA is
expected to have sufficient resolution for such observations,
but probably not sufficient sensitivity.

The minihalos which produce the ``21-cm forest'' of emission lines
described above have appreciable optical depth to 21-cm absorption.
The same minihalos will therefore produce a ``21-cm forest'' of 
{\it absorption\/} lines, too, in the spectra of radio continuum sources if
the latter are discovered at high redshift ($z>6$), prior to the end of
the reionization epoch. We plot illustrative 21-cm absorption line
profiles for absorbing minihalos of different masses at $z_{\rm abs}=9$
in Figure~\ref{Martel_ea}b. These absorption features should be
observable provided sufficiently bright ($\sim1\,\rm mJy$) background
source exists.

\section*{21-CM RADIATION BACKGROUND}

The beam-averaged differential temperature $\overline{\delta T_b}$ 
within a given beam of
angular size $\Delta\theta_{\rm beam}$ is calculated
by integrating our results for individual minihalos over the 
Press-Schechter mass function of minihalos sampled by the beam. We consider the currently-favored flat 
$\Lambda$CDM model ($\Omega_0=0.3$,
$\lambda_0=0.7$, COBE-normalized, $\Omega_bh^2=0.02$, $h=0.7$), with
primordial power spectrum index $n_p=0.9$, 1, and 1.1.

\begin{figure}[b!]
\vspace{-30pt}
  \includegraphics[height=2.8in]{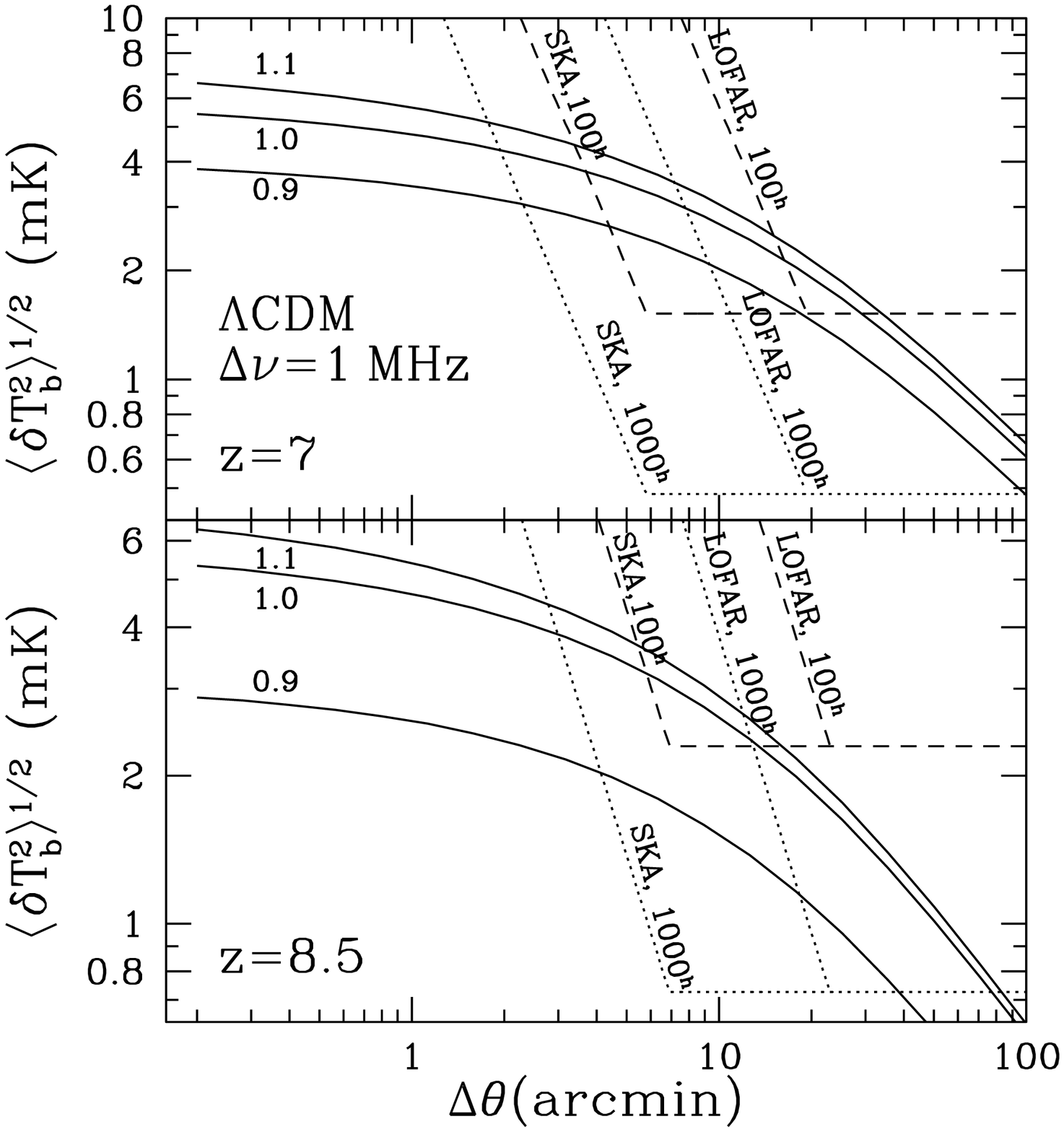}
  \includegraphics[height=2.8in]{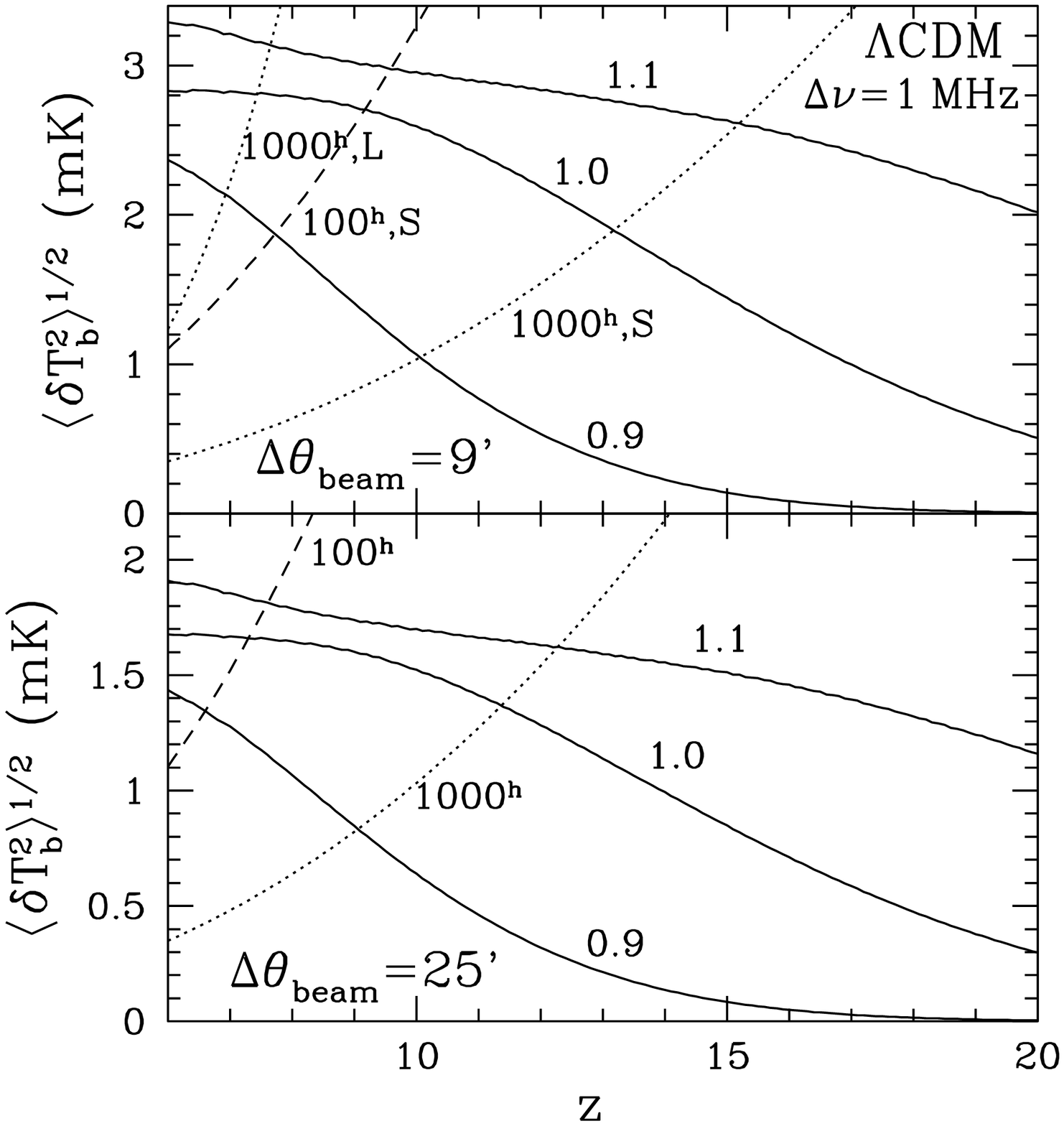}
\vspace{5pt}
\caption{(a) (left) Predicted 3-$\sigma$ differential antenna temperature 
fluctuations at $z=7$ ($\nu_{\rm rec}=177.5\,{\rm MHz}$, top panel) 
and $z=8.5$ ($\nu_{\rm rec}=150\,{\rm MHz}$, bottom panel) for 
bandwidth $\Delta\nu_{\rm obs}=1\,{\rm MHz}$ vs. angular scale 
$\Delta\theta_{\rm beam}$ for $\Lambda$CDM models with tilt $n_p=0.9$, 1.0, 
and 1.1, as labeled (solid curves). Also indicated is the predicted
sensitivity of LOFAR and SKA for a confidence level of 5 times the noise
level after integration times of 100 hr (dashed lines) and 1000 hr
(dotted lines), with compact subaperture (horizontal lines) and extended
configuration needed to achieve higher resolution (diagonal lines).
(b) (right) Predicted 3-$\sigma$ differential antenna temperature 
fluctuations at $\Delta\theta_{\rm beam}=9'$ (top panel) and $25'$ 
(bottom panel). Symbols have same meaning as in the left panels.
Letters ``L'' and ``S'' correspond to LOFAR and SKA, respectively.}
\label{Martel3+4}
\end{figure}

In principle, the variation of $\overline{\delta T_b}$ with observed frequency
should permit a discrimination between the 21-cm emission from minihalos and
the CMB and other backgrounds, due to their very different frequency
dependences. However, the average differential brightness temperature
of the minihalo background is very low and its evolution is fairly smooth,
so such measurement may be difficult in practice with
currently planned instruments like LOFAR and SKA. The angular fluctuations
in this emission, on the other hand, should be much easier to
detect.
The amplitude of $q$-$\sigma$ angular fluctuations (i.e. $q$ times the
rms value) in the differential antenna temperature is given in the linear
regime by $\langle\delta T_b^2\rangle^{1/2}/
\overline{\delta T_b}=qb(z)\sigma_p$,
where $\sigma_p$ is the rms density fluctuation at redshift $z$
in a randomly placed cylinder which corresponds to the observational
volume defined by the detector angular beam size, $\Delta\theta_{\rm beam}$,
and frequency bandwidth, $\Delta\nu_{\rm obs}$, and $b(z)$ is the
bias factor which accounts for the clustering of rare density peaks
relative to the mass.

Illustrative results are plotted for 3-$\sigma$ fluctuations as a function of
$\Delta\theta_{\rm beam}$ for $z=7$ and 8.5, in
Figure~\ref{Martel3+4}a, along with the expected sensitivity limits for the
planned LOFAR (300 m filled aperture) and SKA (1 km filled aperture) arrays.
We plot in Figure~\ref{Martel3+4}b the predicted 
spectral variation of these fluctuations vs. redshift $z$ for illustrative
beam sizes of $\Delta\theta_{\rm beam}=9'$ and $25'$. These 3-$\sigma$
fluctuations should be observable with both LOFAR and SKA with integration
times of between 100 and 1000 hours. For a $25'$ beam, for example,
3-$\sigma$ fluctuations can be detected for untilted $\Lambda$CDM by
both with a 100 hours integration for $z\sim6-7.5$ and a 1000 hours
integration for $z\lesssim11.5$, while for a $9'$ beam, SKA can detect them
after 100 hours for $z\lesssim9$ and after 1000 hours for $z\lesssim13$.

\bibliographystyle{aipproc}

\end{document}